\documentclass[aps,showpacs,prb,twocolumn,superscriptaddress,floatfix,longbibliography]{revtex4-2}
\usepackage[pdftex]{graphicx}
\usepackage{amsbsy, amssymb, amsmath, bm, mathtools}
\usepackage{array}
\usepackage{tabularx}
\usepackage[section]{placeins}
\usepackage[colorlinks=true,linkcolor=blue,citecolor=blue]{hyperref}
\usepackage{url}
\usepackage{enumitem}
\usepackage{color}

\newcolumntype{Y}{>{\raggedright\arraybackslash}X}

\newcolumntype{L}{>{\raggedright\arraybackslash\hspace{6pt}}X<{\hspace{6pt}}}
\begin{document}
\title{Two-dimensional percolation with algebraically decaying interactions II: Critical exponents in the long-range regime}

\author{Ziyu Liu}
\thanks{These two authors contributed equally to this work}
\affiliation{Hefei National Research Center for Physical Sciences at the Microscale and School of Physical Sciences, University of Science and Technology of China, Hefei 230026, China}

\author{Tianning Xiao}
\thanks{These two authors contributed equally to this work}
\affiliation{Hefei National Research Center for Physical Sciences at the Microscale and School of Physical Sciences, University of Science and Technology of China, Hefei 230026, China}

\author{Zhijie Fan}
\email{zfanac@ustc.edu.cn}
\affiliation{Hefei National Research Center for Physical Sciences at the Microscale and School of Physical Sciences, University of Science and Technology of China, Hefei 230026, China}
\affiliation{Hefei National Laboratory, University of Science and Technology of China, Hefei 230088, China}
\affiliation{Shanghai Research Center for Quantum Science and CAS Center for Excellence in Quantum Information and Quantum Physics, University of Science and Technology of China, Shanghai 201315, China}

\author{Youjin Deng}
\email{yjdeng@ustc.edu.cn}
\affiliation{Hefei National Research Center for Physical Sciences at the Microscale and School of Physical Sciences, University of Science and Technology of China, Hefei 230026, China}
\affiliation{Hefei National Laboratory, University of Science and Technology of China, Hefei 230088, China}
\affiliation{Shanghai Research Center for Quantum Science and CAS Center for Excellence in Quantum Information and Quantum Physics, University of Science and Technology of China, Shanghai 201315, China}

\begin{abstract}
We present a comprehensive Monte Carlo study of two-dimensional bond percolation with algebraically decaying connection probabilities $p(r)\propto 1/r^{2+\sigma}$, establishing the universality diagram in the long-range (LR) regime for $\sigma\le2$. Using the event-based ensemble method, we simulate systems with linear sizes up to $L=16384$ and investigate three universality regimes: LR Wilson--Fisher (WF) A ($1<\sigma\le2$), LR Wilson--Fisher B ($2/3<\sigma\le1$), and LR mean-field (MF) ($0<\sigma\le2/3$). In the LR-WF-B regime, the anomalous dimension is consistent with $\eta=2-\sigma$, in agreement with mathematical results for $2/3<\sigma<1$, while the correlation-length exponent $\nu(\sigma)$ exhibits nontrivial, non-Gaussian variation. In the LR-WF-A regime, although $\eta$ remains close to $2-\sigma$ for smaller $\sigma$, statistically resolvable deviations $\delta\eta(\sigma)=\eta-(2-\sigma)>0$ start to appear near $\sigma\simeq3/2$ and grow toward the short-range crossover at $\sigma=2$. Finally, by complementing the event-based simulations with conventional ensemble simulations, we reveal the coexistence of complete-graph asymptotics and LR Gaussian-fixed-point scaling in the LR-MF regime. These results further clarify the critical properties in long-range percolation and provide crucial benchmarks for long-range statistical systems.
\end{abstract}

\maketitle

\section{Introduction}
\label{sec:intro}

Long-range (LR) interactions that decay algebraically with distance provide a direct route to critical behavior beyond conventional short-range (SR) universality~\cite{campa2014, dysonExistencePhasetransitionOnedimensional1969, PhysRev.187.732, fisher1972, cardyOnedimensionalModels11981}. In a $d$-dimensional system with coupling strength or connection probability proportional to $1/r^{d+\sigma}$ (with $r$ the distance between two sites), the exponent $\sigma$ tunes the effective interaction range: small $\sigma$ enhances long-distance connectivity and can drive the system toward mean-field-like behavior, whereas large $\sigma$ suppresses LR effects and restores SR criticality. Between these limits, nontrivial LR fixed points may emerge, leading to critical behavior distinct from both SR and mean-field (MF) universality~\cite{fisher1972, maghrebi2017, xiao_two-dimensional_2024, yao2025nonclassicalregimetwodimensionallongrange}.

The crossover between LR and SR universality is characterized by a threshold $\sigma^*$: for $\sigma<\sigma^*$ the LR tail is relevant and governs the criticality, while for $\sigma>\sigma^*$ the SR fixed point is recovered. This problem was first extensively discussed in the LR-O$(n)$ spin model. Fisher \textit{et al.}~\cite{fisher1972} argued that the LR tail becomes irrelevant at $\sigma=2$, suggesting $\sigma^*=2$ and yielding the Gaussian-form relation for the anomalous dimension $\eta$ (defined through the critical correlation function $g(r) \asymp r^{-d+2-\eta}$):  $\eta=2-\sigma$, valid to $\mathcal O(\epsilon^3)$ with $\epsilon=2\sigma-d$ throughout the LR regime. The resulting discontinuity of exponents at $\sigma^*$ was later resolved by Sak, who proposed a shifted boundary $\sigma^*=2-\eta_{\mathrm{SR}}$~\cite{sak1973}, where 
$\eta_{\mathrm{SR}}$ is the anomalous dimension of the SR universality class. 
Subsequent work explored alternative renormalization group (RG) scenarios, numerical tests,
and the stability of the crossover picture~\cite{
vanenter1982,
YAMAZAKI1977207,
YAMAZAKI1978446,
honkonen1989crossover,
luijtenblote1997,
luijtenblote2002}.
The precise location of the LR--SR crossover has been debated for decades~\cite{picco2012, blanchard2013}. Recent numerical studies on two-dimensional (2D) long-range interacting systems provide strong evidence that the crossover occurs at $\sigma^*=2$ and indicate that $\eta$ need not remain locked to $2-\sigma$ over the entire interval $\sigma<2$~\cite{yao2025nonclassicalregimetwodimensionallongrange, xiao_two-dimensional_2024, yao2025spontaneoussymmetrybreakingtwodimensional}. Field-theoretical calculations have also offered a complementary perspective and are consistent with this scenario~\cite{li4eExpansionLongrange2026,li20266epsilonexpansionlongrangeleeyang}.

For LR-percolation, a field-theoretical description can be formulated via the LR general epidemic process, leading to an effective $\phi^3$ theory with a fractional kinetic term $k^\sigma$, the ordinary SR kinetic term $k^2$, and a cubic interaction $U\sim\phi^3$~\cite{PhysRevB.31.379, janssenLevyflightSpreadingEpidemic1999, linderLongrangeEpidemicSpreading2008, grassberger_sir_2013, grassbergerTwoDimensionalSIREpidemics2013, PhysRevB.17.2956}. A similar controversy regarding the LR--SR boundary also exists in this case. A shifted boundary at $\sigma^*=2 - \eta_{\mathrm{SR}}$ was proposed based on field-theoretical analysis~\cite{linderLongrangeEpidemicSpreading2008, janssenLevyflightSpreadingEpidemic1999}, while numerical results suggest that $\sigma = 2$~\cite{grassbergerTwoDimensionalSIREpidemics2013}. However, for $d \geq 3$, the SR anomalous dimension is negative, so a naive Sak-type shift would place the crossover at $\sigma>2$, motivating alternative scenarios~\cite{PhysRevB.31.379}. In our preceding paper~\cite{liu2025twodimensionalpercolationmodellongrange}, large-scale simulations of 2D LR bond percolation provided strong evidence that the crossover occurs at $\sigma^*=2$, with the boundary point $\sigma=2$ already belonging to the LR universality class, consistent with recent field-theoretical analyses of the $\phi^3$ model based on the $6-\epsilon$ expansion~\cite{li20266epsilonexpansionlongrangeleeyang}.

Recent studies have provided nontrivial insights into the refined structure within the LR universality regime.
Mathematical results establish that, for LR bond percolation in $d \geq 1$, the critical two-point function decays (up to constants) as $g(r)\asymp r^{-d+\sigma}$, for $\sigma < 1$~\cite{hutchcroft2025criticallongrangepercolationii,hutchcroft2024pointwisetwopointfunctionestimates}. 
Note that in 2D, the regime $\sigma \le 2/3$ is already governed by the LR Gaussian fixed point (LR-GFP), for which $\eta = 2-\sigma$. The result therefore implies that the same Gaussian-form relation remains valid throughout $2/3< \sigma <1$, a regime governed by nontrivial LR fixed points. This motivates $\sigma = 1$ as a special endpoint within the LR universality regime. A complementary perspective comes from the long-range simple random walk (LR-SRW) with an algebraic step-length kernel~\cite{xiao2025universalitydiagramphasetransitions, yao2025spontaneoussymmetrybreakingtwodimensional}. The Green's function of the LR-SRW gives the covariance of the associated long-range Gaussian free field, while its transport behavior probes the effective connectivity of the underlying long-range network~\cite{sheffield2007gaussianfreefields,barlow2009parabolicharnack}. The divergence of the second moment at $\sigma=2$ separates diffusive and superdiffusive transport, while the divergence of the first moment at $\sigma=1$ separates sub-ballistic and hyper-ballistic spreading~\cite{xiao2025universalitydiagramphasetransitions}. The change of transport properties reflects qualitative changes in the geometry of the underlying graph induced by the strong nonlocal connectivity, highlighting the possibility that $\sigma = 1$ is a special point. 

\begin{figure}[t]
\centering
\includegraphics[width=1\linewidth]{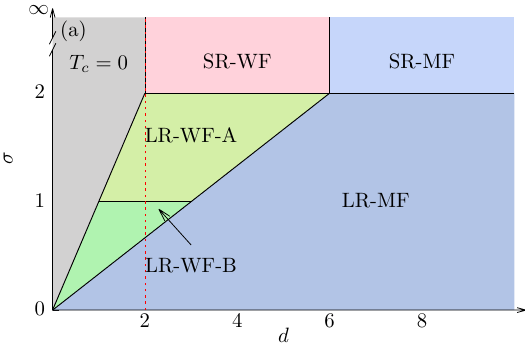}
\includegraphics[width=1\linewidth]{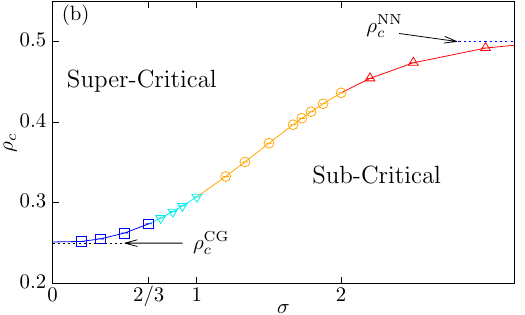}
\caption{
(a) Proposed universality class diagram of the LR-percolation model in the $(d,\sigma)$ plane. ``SR'' and ``LR'' denote short-range and long-range, respectively; ``WF'' represents the Wilson--Fisher fixed point, and ``MF'' denotes mean-field behavior. In the ``$T_c=0$'' regime, percolation occurs only at the trivial fully occupied-bond limit, analogous to a zero-temperature transition in the spin system. The vertical dashed line marks $d=2$, the case studied in this work.
(b) The phase diagram of the 2D LR bond percolation model. 
The symbols mark the estimated critical parameter $\rho_c(\sigma)$ (the number of occupied bonds per site is $2\rho_c$, see the main text) in the different universality regimes: blue squares for LR-MF ($0<\sigma\le2/3$), cyan downward triangles for LR-WF-B ($2/3<\sigma\le1$), orange circles for LR-WF-A ($1<\sigma\le2$), and red upward triangles for SR-WF ($\sigma>2$).
The critical line $\rho_c(\sigma)$ separates the subcritical phase at $\rho<\rho_c$ from the supercritical (percolating) phase at $\rho>\rho_c$. 
The dotted horizontal lines indicate the complete-graph (CG) reference value $\rho_c^{\mathrm{CG}}=1/4$ and the nearest-neighbor (NN) reference value $\rho_c^{\mathrm{NN}}=1/2$.
}
\label{fig:PDCP}
\end{figure}

Based on recent theoretical and numerical progress in LR systems, Ref.~\cite{xiao2025universalitydiagramphasetransitions} proposed a universality class diagram for LR bond percolation in the $(d,\sigma)$ plane, shown in Fig.~\ref{fig:PDCP}(a). The diagram is organized by several boundaries: (i) $\sigma=2$ separates SR and LR universality; (ii) $\sigma=\min(d/3,2)$ marks the LR upper-critical dimension boundary, below which the critical behavior is governed by the LR-GFP (MF theory)~\cite{heydenreich2008meanfield}; (iii) $d = \min(\sigma,2)$ marks the lower-critical dimension below which no phase transition occurs; and (iv) $\sigma=1$ further subdivides the LR Wilson--Fisher regime into LR-WF-A and LR-WF-B, motivated by the endpoint of the interval for $\eta=2-\sigma$, the transport crossover of the associated LR-SRW, and the possibility that $\eta$ is renormalized only in part of the LR-WF regime.

In this work, we focus on the $d=2$ case, for which the universality diagram contains four regimes [Fig.~\ref{fig:PDCP}(b)].
For $\sigma>2$ (SR-WF), the LR tail is irrelevant and ordinary 2D percolation is recovered~\cite{nienhuis1982coulombgas, staufferaharony1994}. The correlation-length exponent $\nu$ and the anomalous dimension $\eta$ are then given by
\begin{equation}
    \nu=\nu_{\mathrm{SR}}=\frac{4}{3},
    \qquad
    \eta=\eta_{\mathrm{SR}}=\frac{5}{24}.
\end{equation}
For $1<\sigma\le 2$ (LR-WF-A), criticality is governed by a nontrivial LR Wilson--Fisher fixed point~\cite{liu2025twodimensionalpercolationmodellongrange,
li20266epsilonexpansionlongrangeleeyang}, and both $\nu(\sigma)$ and $\eta(\sigma)$ may vary with $\sigma$. For $2/3<\sigma \le 1$ (LR-WF-B), $\eta$ retains its Gaussian-form value $\eta=2-\sigma$~\cite{hutchcroft2025criticallongrangepercolationii, hutchcroft2024pointwisetwopointfunctionestimates}, while $\nu$ remains nontrivial.

For $0<\sigma\le 2/3$ (LR-MF), the transition is governed by the LR-GFP with $\eta^{\mathrm{GFP}}=2-\sigma$ and $\nu^{\mathrm{GFP}}=1/\sigma$~\cite{hutchcroft2025criticallongrangepercolationi,hutchcroft2025criticallongrangepercolationiii, xiao2025universalitydiagramphasetransitions}. However, in analogy with the SR model above the upper critical
dimension, the finite-size scaling (FSS) behavior of the model can
display \emph{two} coexisting scalings: a spatially dependent Gaussian
fixed point contribution and a distance-independent complete-graph (CG)
contribution~\cite{
hutchcroft2023torusplateau,
fangGeometricUpperCritical2022,
fangGeometricScalingBehaviors2023,
PhysRevE.110.044140}.
Concretely, the critical two-point function admits the two-scale form
\begin{equation}
    g(r,L)
    \sim
    r^{-(d-2+\eta^{\mathrm{GFP}})}\,\widetilde{g}(r/L)
    +
    c_1 L^{-(d-2+\eta_L)},
    \label{eq:gM1}
\end{equation}
where the first term captures the LR-GFP scaling in real space, while the second term produces an $L$-dependent plateau characteristic of CG asymptotics. As a result, global/zero-momentum observables are controlled by CG scaling exponents $(\nu_L,\eta_L)=(3/d,2-d/3)$, whereas spatial correlations and nonzero-momentum observables suppress the plateau and directly probe the LR-GFP exponents $(\nu^{\mathrm{GFP}},\eta^{\mathrm{GFP}})=(1/\sigma,2-\sigma)$. For $d=2$, the corresponding CG scaling exponents are $\nu_L=3/2$ and $\eta_L=4/3$~\cite{xiao2025universalitydiagramphasetransitions}.

\begin{figure}[t]
\centering
\includegraphics[width=\linewidth]{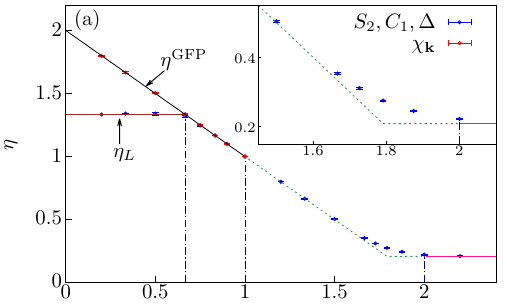}
\includegraphics[width=\linewidth]{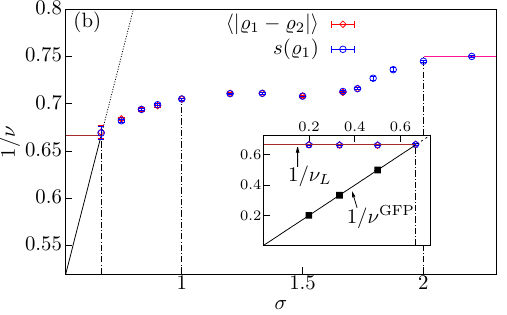}
\caption{
Overview of the critical exponents across the two-dimensional LR-percolation regimes.
(a) Anomalous dimension $\eta$ as a function of $\sigma$, obtained from $S_2$, $C_1$, $\Delta$, and $\chi_{\mathbf{k}}$, defined in Sec.~\ref{sec:model} B.
The dashed line denotes the Sak reference $\eta=\max(2-\sigma,\eta_{\mathrm{SR}})$~\cite{sak1973}, and the horizontal line marks the SR value.
For $2/3<\sigma\le1$, the data are consistent with $\eta=2-\sigma$. In the LR-WF-A regime, deviations from this form become visible near $\sigma\simeq3/2$ and are clearly resolved for larger $\sigma$, while the boundary point $\sigma=2$ remains distinct from the SR value (inset).
For $0<\sigma \le 2/3$, $S_2$, $C_1$, $\Delta$ follow the CG-asymptotic scaling exponent $\eta_L=4/3$, whereas $\chi_{\mathbf{k}}$ follows the GFP scaling $\eta^{\mathrm{GFP}}=2-\sigma$.
(b) Thermal scaling exponent $y_t = 1/\nu$ obtained from $s(\varrho_1)$ and $\langle|\varrho_1-\varrho_2|\rangle$.
In the LR-WF-A and LR-WF-B regimes, $y_t = 1/\nu$ exhibits a nonmonotonic $\sigma$-dependent behavior and differs from both the GFP prediction $y_t^{\mathrm{GFP}}=\sigma$ and the SR value $1/\nu_{\mathrm{SR}} =3/4$.
For $0<\sigma\le2/3$, the event-based estimates approach the CG-asymptotic value $y_t^{\mathrm{CG}}=1/\nu_L=2/3$, while the nonzero-momentum data collapse is consistent with the GFP exponent $y_t^{\mathrm{GFP}}=1/\nu^{\mathrm{GFP}}=\sigma$ (inset).
}
\label{fig:exponent_overview}
\end{figure}

We perform large-scale Monte Carlo simulations utilizing the event-based ensemble method~\cite{fan_universal_2020, liExplosivePercolationObeys2023, li_explosive_2024, shi_universality_2025}, reaching linear sizes up to $L=16384$. The estimated critical exponents are summarized in Fig.~\ref{fig:exponent_overview} and Tables~\ref{table:rho_c_y_t_eta} and~\ref{table:MF}. In the LR-WF-B regime, our data are consistent with $\eta=2-\sigma$. In the LR-WF-A regime, the lower portion remains compatible with $\eta=2-\sigma$ within numerical uncertainty; however, deviations become visible near $\sigma\simeq3/2$ and are clearly resolved for larger $\sigma$, suggesting a nontrivial anomalous dimension in the LR-WF-A regime. Furthermore, the correlation-length exponent $\nu$ varies nontrivially across the entire LR-WF-A and LR-WF-B regimes, and deviates from the LR-GFP value $\nu^{\mathrm{GFP}}=1/\sigma$. Complementing conventional ensemble methods, we demonstrate that in the LR-MF regime, global observables are governed by the CG finite-size scaling characterized by $\nu_L$ and $\eta_L$, whereas nonzero-momentum observables probe the LR-GFP scaling characterized by $\nu^{\mathrm{GFP}}$ and $\eta^{\mathrm{GFP}}$. Together with the preceding work on the LR--SR crossover~\cite{liu2025twodimensionalpercolationmodellongrange}, these results are in line with the proposed universality class diagram.

The remainder of this paper is organized as follows. 
Section~\ref{sec:model} defines the model, algorithms, and observables.
Section~\ref{sec:two_regime_fss} summarizes the two-scale finite-size scaling form in the LR-MF regime.
Section~\ref{sec:results} presents the numerical results for the LR-WF-A, LR-WF-B, and LR-MF regimes. 
Finally, Sec.~\ref{sec:conclusion} summarizes the conclusions and discusses the remaining open questions.

\begin{table}[t]
\caption{Summary of the percolation thresholds $\rho_c$, correlation-length exponent $\nu$, and anomalous dimension $\eta$ in the SR, LR-WF-A, and LR-WF-B regimes for different values of $\sigma$.}\label{table:rho_c_y_t_eta}
\renewcommand{\arraystretch}{1.1}
\begin{tabularx}{0.87\columnwidth}{l|lll}
\hline\hline
\; $\sigma$\, &\quad $\rho_c$       & \qquad $1/\nu$    & \qquad $\eta$ \\ \hline
\; NN\,       &\quad 0.5            & \qquad 3/4        & \qquad 5/24            \\
\; 11/5\,     &\quad 0.454\,408(2)  & \qquad 0.750(1)   & \qquad 0.210(2)        \\
\hline
\; 2\,        &\quad 0.436\,504(3)  & \qquad 0.745(2)   & \qquad 0.223(2)        \\
\; 15/8\,     &\quad 0.423\,004(4)  & \qquad 0.736(2)   & \qquad 0.245(2)        \\
\; 43/24\,    &\quad 0.413\,053(3)  & \qquad 0.725(3)   & \qquad 0.275(2)        \\
\; 19/11\,    &\quad 0.404\,897(3)  & \qquad 0.716(1)   & \qquad 0.311(3)        \\
\; 5/3\,      &\quad 0.396\,913(3)  & \qquad 0.712(1)   & \qquad 0.353(3)        \\
\; 3/2\,      &\quad 0.373\,996(3)   & \qquad 0.708(1)   & \qquad 0.504(3)        \\
\; {4/3}\,      &\quad {0.350\,764(4)}  & \qquad 0.711(1)   & \qquad 0.666(2)        \\
\; 6/5\,      &\quad 0.332\,728(3)   & \qquad 0.711(1)   & \qquad 0.800\,0(5)     \\
\hline
\; 1\,        &\quad 0.307\,591(4)   & \qquad 0.705(1)   & \qquad 1.000\,0(1)        \\
\; 9/10\,     &\quad 0.296\,216(6)   & \qquad 0.699(1)   & \qquad 1.100\,0(1)        \\
\; 5/6\,      &\quad 0.289\,187(5)  & \qquad 0.694(2)   & \qquad 1.167(2)        \\
\; 3/4\,      &\quad 0.281\,123(6)  & \qquad 0.683(2)   & \qquad 1.246(5)        \\
\hline
\; {2/3}\,      &\quad {0.273\,94(1)}   & \qquad {0.670(7)}   & \qquad {1.32(1)}        \\ \hline\hline
\end{tabularx}
\vspace{1em}
\caption{
Summary of the percolation thresholds $\rho_c$ and exponent estimates in the LR-MF regime. The fitted values of $1/\nu_L$, $\eta_L$, and $\eta^{\mathrm{GFP}}$ are consistent with the respective predictions $2/3$, $4/3$, and $2-\sigma$; the $1/\nu^{\mathrm{GFP}}$ column lists the GFP prediction $1/\nu^{\mathrm{GFP}}=\sigma$.
}
\label{table:MF}
\renewcommand{\arraystretch}{1.2}
\begin{tabularx}{\columnwidth}{l|lllll}
\hline\hline
$\sigma$ & $\rho_c$      & \quad$1/\nu_{L}$ & \quad$1/\nu^{\mathrm{GFP}}$ & \;$\eta_{L}$ & \quad$\eta^{\mathrm{GFP}}$ \\
\hline
1/2      & {0.262\,63(2)}  & \quad0.665(2)              & \quad1/2                    & \;1.34(1)             & \quad1.504(5)              \\
1/3      &  {0.255\,4(1)}             & \quad0.666(1)              & \quad1/3                    & \;1.340(5)             & \quad1.666(9)               \\
1/5      & {0.252\,1(1)}     & \quad0.666(1)              & \quad1/5                    & \;1.333(2)             & \quad1.796(5)              \\ \hline\hline
\end{tabularx}
\end{table}

\section{Model, Algorithms, and Observables}
\label{sec:model}

\subsection{Model}

We consider 2D LR bond percolation on an $L\times L$ square lattice with periodic boundary conditions. For two distinct sites $i$ and $j$, the bond percolation probability depends on their minimum-image distance $r_{ij}$ as~\cite{frenkel_understanding_2002,christiansen_phase_2019, agrawal_kinetics_2021}
\begin{equation}
    p_{ij}(\rho,\sigma,L) = \rho\,\frac{C(\sigma,L)}{r_{ij}^{2+\sigma}},
    \label{eq:normalized_percolation_prob}
\end{equation}
where $\rho$ is the normalized bond parameter. The normalization factor $C(\sigma,L)$ is defined by
\begin{equation}
   C(\sigma,L) \sum_{j\ne i}\frac{1}{r_{ij}^{2+\sigma}}=4 ,
\end{equation}
so that $\sum_{j\ne i}p_{ij}=4\rho$. 
With this convention, in the $\sigma \to \infty$ limit, the model reduces to the square lattice bond percolation model with $C=1$ and the critical bond probability $\rho_c=1/2$. In the opposite limit,  $\sigma\to -2$, the normalized connection probability becomes distance independent, and the model approaches the CG limit of random-graph percolation; one has $C = 4/(L^2-1)$ and the critical point $\rho_c=1/4$.
This normalization, following Ref.~\cite{liu2025twodimensionalpercolationmodellongrange}, ensures that the expected number of occupied bonds per lattice site is $2\rho$, independently of $\sigma$ and $L$.

\subsection{Algorithms and Observables}

We perform extensive Monte Carlo simulations using the event-based ensemble (EB) method, supplemented by the conventional ensemble (CE) at a fixed bond parameter. The EB offers several advantages for extracting critical properties compared with conventional percolation simulations. By identifying a pseudo-critical point in each realization, it enables accurate estimation of critical exponents without precise prior knowledge of the percolation threshold. It also exhibits clean FSS behavior, thereby providing statistically robust access to universal quantities. The CE is used as a complementary method to identify the LR-GFP scaling contribution in the LR-MF regime.

Because the algorithmic details are identical to those in Ref.~\cite{liu2025twodimensionalpercolationmodellongrange}, they are only briefly reviewed below.

\textit{Event-based ensemble}. In the EB, the percolation process is modeled as a sequential bond-insertion process~\cite{newmanziff2001, fan_universal_2020, liExplosivePercolationObeys2023, li_explosive_2024, shi_universality_2025}. Let $\mathcal{C}_1(t)$ denote the size of the largest cluster after the $t$-th bond insertion. The gap at step $t$ is defined as $\delta(t) = \mathcal{C}_1(t) - \mathcal{C}_1(t-1)$. The first pseudo-critical event is defined as the step $t_1$ at which $\delta(t)$ reaches its largest value during the bond-insertion process. The corresponding gap is denoted by $\delta_1$, and the normalized bond parameter is $\varrho_1 = t_1/(2N)$, where $N = L^2$. We also record the density $\varrho_2$ associated with the second pseudo-critical event, corresponding to the second-largest gap value. The configuration at $t_1$ can be reconstructed, and clusters can then be identified by breadth-first search (BFS), with various observables measured on that configuration.

During each percolation process, we record the following quantities.
\begin{enumerate}[
  label=(\alph*),        
  leftmargin=2em,         
  labelwidth=1.5em,     
  labelsep=0.5em,        
  align=left          
]

\item the positions of the first and second pseudo-critical points, $\varrho_1$ and $\varrho_2$. 
\item the gap value at the first pseudo-critical point $\delta_1$~\cite{liu2025twodimensionalpercolationmodellongrange}. 
\end{enumerate}

For the percolation configuration at the first pseudo-critical point, we measure the following quantities.
\begin{enumerate}[
  label=(\alph*),        
  leftmargin=2em,         
  labelwidth=1.5em,     
  labelsep=0.5em,        
  align=left          
]

\item The size of the largest cluster $\mathcal{C}_1$.

\item The second moment of the cluster-size distribution, $\mathcal{S}_2=\sum_i \mathcal{C}_i^2$.

\item The Fourier mode of the magnetization $\mathcal{M}_{\mathbf{k}}=\bigl|\sum_i s_i\,e^{i\,\mathbf{k}\cdot\mathbf{r}_i}\bigr|$, obtained by assigning a random sign $\epsilon_C \in \{\pm 1\}$ to each cluster and setting $s_i = \epsilon_{C}$ for every site $i$ belonging to the cluster. Here, $\mathbf{k}=(2\pi/L,0)$ is the smallest nonzero wavevector along the $x$ direction.

\end{enumerate}

By taking the ensemble average $\langle \cdot \rangle$ of these measurements, we obtain the following observables:

\begin{enumerate}[
  label=(\alph*),        
  leftmargin=2em,         
  labelwidth=1.5em,     
  labelsep=0.5em,        
  align=left          
]

\item The averaged pseudo-critical gap $\Delta = \langle \delta_{1} \rangle$.

\item The averaged pseudo-critical point $\rho_1 = \langle \varrho_1 \rangle$ and $\rho_2 = \langle \varrho_2 \rangle$.

\item The standard deviation of the first pseudo-critical point $s(\varrho_1) = \sqrt{\langle \varrho_1^2 \rangle - \langle \varrho_1 \rangle^2}$.

\item The critical window $\langle |\varrho_1 - \varrho_2| \rangle$.

\item The averaged largest-cluster size $C_1=\langle\mathcal{C}_1\rangle$.

\item The susceptibility $S_2=L^{-2}\langle\mathcal{S}_2\rangle$.

\item The Fourier-mode susceptibility $\chi_{\mathbf{k}}=L^{-2}\langle\mathcal{M}_{\mathbf{k}}^2\rangle$.

\end{enumerate}

\textit{Conventional ensemble}. In the CE, configurations are generated at a fixed parameter $\rho$ using the accelerated bond-generation scheme described in Ref.~\cite{liu2025twodimensionalpercolationmodellongrange}, in which the next activated bond for each displacement type is sampled directly. After the bonds are generated, clusters are identified using BFS.

In the CE, the same cluster observables $C_1(\rho,L)$, $S_2(\rho,L)$, and $\chi_{\mathbf{k}}(\rho,L)$ can be measured at fixed $\rho$. In the present work, CE data are used primarily for the fixed-density analysis of $\chi_{\mathbf{k}}$ in the LR-MF regime.

\section{Finite-size scaling in the LR-MF regime}
\label{sec:two_regime_fss}

The FSS in the SR-WF and LR-WF regimes is governed by a single set of exponents~\cite{liu2025twodimensionalpercolationmodellongrange}. In the LR-MF regime ($0<\sigma<d/3$), however, finite-size criticality can exhibit two coexisting scalings, analogous to SR models above their upper critical dimension~\cite{brezin1982fss, binderNauenbergPrivmanYoung1985, Wittmann2014, PhysRevE.110.044140, PhysRevE.109.034125, PhysRevE.104.064108, PhysRevE.102.022125, xiao2025universalitydiagramphasetransitions}.

Let $t=(\rho-\rho_c)/\rho_c$. We assume that the singular part of the FSS functional takes the two-scale form
\begin{equation}
\begin{split}
    f_s(t,h,L)
    =
    L^{-d}
    \tilde f_0(tL^{y_t^{\mathrm{GFP}}},hL^{y_h^{\mathrm{GFP}}})
    \\+
    L^{-d}
    \tilde f_1(tL^{y_t^{\mathrm{CG}}},hL^{y_h^{\mathrm{CG}}}) ,
\end{split}
\label{eq:feM}
\end{equation}
where $h$ denotes the magnetic scaling field, and $\tilde f_0$ and $\tilde f_1$ are scaling functions. For the two scaling contributions, the thermal and magnetic exponents are related to the corresponding correlation-length exponent and anomalous dimension by
\begin{equation}
    y_t=\frac{1}{\nu},
    \qquad
    y_h=\frac{d+2-\eta}{2}.
    \label{eq:scaling_dimension_relations}
\end{equation}
Substituting the LR-GFP exponents $(\nu^{\mathrm{GFP}},\eta^{\mathrm{GFP}})=(1/\sigma,2-\sigma)$ and the CG scaling exponents $(\nu_L,\eta_L)=(3/d,2-d/3)$ into Eq.~\eqref{eq:scaling_dimension_relations}, we obtain
\begin{align}
    & y_t^{\mathrm{GFP}} = \sigma, 
    && y_h^{\mathrm{GFP}} = \frac{d+\sigma}{2},
    \nonumber \\
    & y_t^{\mathrm{CG}} = \frac{d}{3}, 
    && y_h^{\mathrm{CG}} = \frac{2d}{3}.
    \label{eq:GFP_CG_exponents}
\end{align}
The associated two-point function can be written as
\begin{equation}
    g(r,L)
    \sim
    r^{2y_h^{\mathrm{GFP}}-2d}\,\tilde g(r/L)
    +
    c_1 L^{2y_h^{\mathrm{CG}}-2d},
    \label{eq:gM2}
\end{equation}
with a distance-dependent LR-GFP part and an $L$-dependent CG plateau.

As a consequence, zero-momentum/global observables are CG-dominated,
\begin{equation}
    S_2 \sim L^{2y_h^{\mathrm{CG}}-d},
    \qquad
    C_1,\Delta \sim L^{y_h^{\mathrm{CG}}},
    \label{eq:S2_CG_scaling}
\end{equation}
while nonzero-momentum susceptibilities probe the LR-GFP contribution,
\begin{equation}
    \chi_{\mathbf{k}}
    \sim
    L^{2y_h^{\mathrm{GFP}}-d}.
    \label{eq:chik_GFP_scaling}
\end{equation}
For 2D LR-percolation, this reduces to
\begin{equation}
    S_2\sim L^{2/3},
    \qquad
    C_1,\Delta\sim L^{4/3},
    \qquad
    \chi_{\mathbf{k}}\sim L^\sigma .
    \label{eq:SLR_magnetic_scaling_2d}
\end{equation}

Thermally, EB pseudo-critical fluctuations probe the CG exponent,
\begin{equation}
    s(\varrho_1),\ \langle|\varrho_1-\varrho_2|\rangle
    \sim
    L^{-y_t^{\mathrm{CG}}},
    \qquad
    y_t^{\mathrm{CG}}=\frac{1}{\nu_L},
    \label{eq:yt_CG}
\end{equation}
whereas the LR-GFP prediction $y_t^{\mathrm{GFP}}=\sigma$ is tested by a fixed-density CE collapse of $\chi_{\mathbf{k}}$,
\begin{equation}
    \chi_{\mathbf{k}}(\rho,L)
    =
    L^{2y_h^{\mathrm{GFP}}-d}
    \tilde\chi_{\mathbf{k}}\left(tL^{y_t^{\mathrm{GFP}}}\right)
    +c_0.
    \label{eq:chi_k}
\end{equation}

\section{Results}
\label{sec:results}

We now test the universality regimes proposed in Fig.~\ref{fig:PDCP}. The preceding paper focused on the vicinity of the LR--SR crossover at $\sigma=2$~\cite{liu2025twodimensionalpercolationmodellongrange}; here we focus on the LR side, especially the LR-WF-A, LR-WF-B, and LR-MF regimes. In the two interacting LR-WF regimes, we study $\sigma=19/11$, $5/3$, $3/2$, $4/3$, $6/5$, $1$, $9/10$, $5/6$, and $3/4$. We also estimate the boundary-point behavior at $\sigma=2/3$ and perform the two-scale LR-MF analysis at $\sigma=1/2$, $1/3$, and $1/5$. We use EB simulations to determine the critical thresholds and study FSS at criticality, and CE simulations to probe the nonzero-momentum susceptibility in the LR-MF regime. The largest EB simulations reach $L=16384$, and each data point is based on more than $10^6$ samples. Statistical errors are estimated using the binning method. The main exponent estimates are summarized in Tables~\ref{table:rho_c_y_t_eta} and~\ref{table:MF}. Data for $\sigma\ge43/24$ are taken from Ref.~\cite{liu2025twodimensionalpercolationmodellongrange} and included for completeness.

\subsection{Percolation thresholds $\rho_c$ and correlation-length exponent $\nu$}
\label{sec:nu}

We determine the percolation thresholds $\rho_c$ and the correlation-length exponent $\nu = 1/y_t$ in the LR-WF-A and LR-WF-B regimes. 
As stated in the previous section, the EB method provides an efficient way to estimate critical exponents, such as $\nu$ and $\eta$, without precise prior knowledge of $\rho_c$. 
At each $\sigma$, we simulate systems of various sizes using the EB method and estimate $\nu$ using $s(\varrho_1)$ and $\langle |\varrho_1 - \varrho_2| \rangle$. These two observables, as a function of $L$, are fitted using the following ansatz~\cite{li_explosive_2024, liu2025twodimensionalpercolationmodellongrange}
\begin{align}
    \mathcal{O}(L) = L^{-1/\nu}(a+b_1L^{-y_1}+b_2L^{-y_2}),
    \label{eq:s1}
\end{align}
where $a$, $b_1$, and $b_2$ are nonuniversal constants, and $y_1$ and $y_2$ are correction exponents. 
Using this fit, we obtain a high-precision estimate of $1/\nu$. We then substitute this value into the fitting ansatz for the first pseudo-critical point $\rho_1$ to obtain an accurate estimate of the critical threshold $\rho_c$,
\begin{align}
    \rho_1=\rho_c +L^{-1/\nu}(a+b_1L^{-y_1}).
    \label{eq:rho1}
\end{align}
The detailed fitting procedures and fit tables for $s(\varrho_1)$ are provided in Appendix~\ref{sec:fitting_details}. Both $s(\varrho_1)$ and $\langle |\varrho_1 - \varrho_2| \rangle$ yield highly consistent results. The final estimates of $\rho_c$ and $1/\nu$ are summarized in Table~\ref{table:rho_c_y_t_eta} for the LR-WF-A and LR-WF-B regimes.
The orange and cyan points in Fig.~\ref{fig:PDCP}(b) mark the critical thresholds in these two regimes.

Fig.~\ref{fig:exponent_overview}(b) shows estimates of $1/\nu$ obtained from fits of $s(\varrho_1)$ and $\langle |\varrho_1-\varrho_2|\rangle$ as a function of $\sigma$. In both interacting LR-WF regimes, the estimates remain well below the GFP prediction $y_t=1/\nu=\sigma$ (dashed line).
Instead, the thermal exponent $y_t=1/\nu$ shows a nontrivial, nonmonotonic dependence on $\sigma$. Starting near the LR-MF value at the lower boundary, $y_t$ increases and reaches a local maximum between $\sigma=6/5$ and $4/3$. It then decreases to a shallow minimum near $\sigma\simeq3/2$ and approaches the SR value as $\sigma$ approaches $2$. The shallow structure near $\sigma\simeq3/2$ may reflect competition among the LR and SR kinetic contributions, $k^\sigma$ and $k^2$, and the interaction term $U$ in the effective-Hamiltonian picture. Overall, $y_t$ is $\sigma$ dependent in both interacting LR regimes and differs from the GFP value.

\subsection{Anomalous dimension $\eta$ in the LR-WF-A and LR-WF-B regimes}

In the conventional RG framework~\cite{fisher1972,sak1973,linderLongrangeEpidemicSpreading2008}, the nonanalytic term $k^\sigma$ in the continuum Hamiltonian is argued not to be renormalized by the analytic interaction term. This argument leads to the expectation that $\eta=2-\sigma$ is exact throughout the nonclassical interval $2/3<\sigma<\sigma^*$. Within this framework, the relation is often taken to be model-independent and forms the basis of several theoretical studies~\cite{sak1973,linderLongrangeEpidemicSpreading2008}.
Mathematical results for 2D LR-percolation establish $\eta=2-\sigma$ in the open interval $2/3<\sigma<1$~\cite{hutchcroft2025criticallongrangepercolationii,hutchcroft2024pointwisetwopointfunctionestimates}. For LR-O$(n)$ $|\varphi|^4$ models in $d=1,2,3$ (including the $n=0$ weakly self-avoiding walk), Ref.~\cite{lohmannCriticalTwopointFunction2017} established the same relation for $\sigma=(d+\epsilon)/2$ with sufficiently small $\epsilon>0$, which seemingly implies that $\eta(\sigma)$ is infinitely differentiable at $\sigma=d/2$.

Nevertheless, numerical studies of LR-O$(n)$ models and LR-percolation indicate that $\eta$ can deviate from the Gaussian-form value near the LR--SR crossover~\cite{xiao_two-dimensional_2024,yao2025nonclassicalregimetwodimensionallongrange,liu2025twodimensionalpercolationmodellongrange}. Such a deviation is also predicted by recent field-theoretical studies of the LR-O$(n)$ model and LR-percolation, based on a $4-\epsilon$ expansion and a $6-\epsilon$ expansion near $\sigma=2$~\cite{li4eExpansionLongrange2026,li20266epsilonexpansionlongrangeleeyang}.

We therefore determine $\eta$ numerically in the LR-WF-A and LR-WF-B regimes to examine the range of validity of the GFP relation $\eta=2-\sigma$.

In the EB simulation, the anomalous dimension $\eta$ can be extracted from the scaling of $C_1$, $S_2$, and $\Delta$ at the pseudo-critical point. The asymptotic scaling of $S_2$ is fitted using
\begin{align}
\mathcal{O} = L^{2 - \eta}(a + b_1 L^{-y_1} + b_2 L^{-y_2}) +c_0,
\label{eq:S2_scaling}
\end{align}
while $C_1$ and $\Delta$ are fitted using
\begin{align}\label{eq:Delta_scaling}
\mathcal{O} = L^{2 - {\eta}/{2}}(a + b_1 L^{-y_1}+ b_2 L^{-y_2}) + c_0,
\end{align}
where $y_1$ and $y_2$ are finite-size correction exponents, $a$, $b_1$, and $b_2$ are nonuniversal constants, and $c_0$ is a background term. The fitting results for $C_1$ and $S_2$ are consistent with those for $\Delta$; we therefore present only the detailed fits for $\Delta$ in Appendix~\ref{sec:fitting_details} as a representative case. The final estimates of $\eta$ summarized in Table~\ref{table:rho_c_y_t_eta} take into account fits of all three observables.

Fig.~\ref{fig:exponent_overview}(a) shows the estimated values of $\eta$ as a function of $\sigma$. The green dashed line marks Sak's criterion, $\eta=\max(2-\sigma,\eta_{\mathrm{SR}})$, or equivalently $y_h=\max(1+\sigma/2,y_h^{\mathrm{SR}})$. In the LR-WF-B regime, $2/3<\sigma \le 1$, the estimates are consistent with the Gaussian-form relation $\eta=2-\sigma$ within numerical uncertainty. This agreement extends over a narrow interval above $\sigma=1$. A deviation becomes visible near $\sigma\simeq3/2$ and is clearly resolved at larger $\sigma$, as shown in the inset of Fig.~\ref{fig:exponent_overview}(a).

\begin{figure}[t]
    \centering
    \includegraphics[width=1\linewidth]{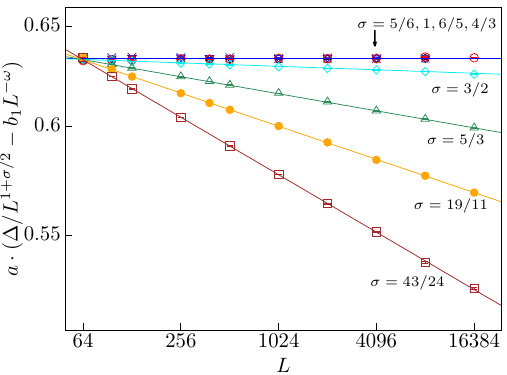}
    \caption{Log-log plot of $\Delta/L^{1+\sigma/2}$ as a function of system size $L$ for various $\sigma$. The fitted leading correction has been subtracted using $\omega(43/24)=0.4231$, $\omega(19/11)=0.4081$, $\omega(5/3)=0.3998$, $\omega(3/2)=0.512$, $\omega(4/3)=0.5497$, $\omega(1)=0.77$, and $\omega(5/6)=0.5867$, where $\omega \equiv y_1$
 denotes the fitted leading correction exponent. Because the $y_1$ term cannot be resolved reliably for $\sigma=6/5$ (see Table~\ref{table:Delta}), no leading-correction subtraction is applied to that curve. A constant normalization is applied to facilitate comparison. A visible slope develops near $\sigma\simeq3/2$ and becomes more pronounced at larger $\sigma$. For $\sigma\le4/3$, the data tend toward the blue guide line.}
    \label{fig:Delta_modified}
\end{figure}

The deviation can be visualized more directly by comparing the critical scaling of $C_1$, $S_2$, and $\Delta$ with the GFP relation. Under the hypothesis $\eta=2-\sigma$, these observables scale as
\begin{align}
S_2 \sim L^{\sigma}, \quad
C_1 \sim L^{1+\sigma/2}, \quad \Delta \sim L^{1+\sigma/2}.
\end{align}
Dividing each observable by its corresponding GFP power law therefore gives a slope that directly probes $\delta\eta=\eta-(2-\sigma)$. 
Fig.~\ref{fig:Delta_modified} shows $\Delta/L^{1+\sigma/2}$ with the leading correction subtracted. To facilitate comparison at large sizes, each curve is multiplied by a constant chosen so that the data points overlap at $L=64$. The curves for $\sigma < 3/2$ collapse onto one horizontal line, indicating consistency with $\eta=2-\sigma$, while for larger values of $\sigma$, the corresponding curves exhibit a clear deviation from the GFP scaling of $\eta$.

Taken together, our results show that the anomalous dimension $\eta$ is consistent with $2-\sigma$ throughout $2/3<\sigma\le1$. 
For $1<\sigma< 3/2$, the data remain compatible with this relation within the present numerical resolution, yet a deviation becomes visible near $\sigma\simeq3/2$ and increases toward the LR--SR crossover.

\begin{figure}[t]
    \centering
    \includegraphics[width=\linewidth]{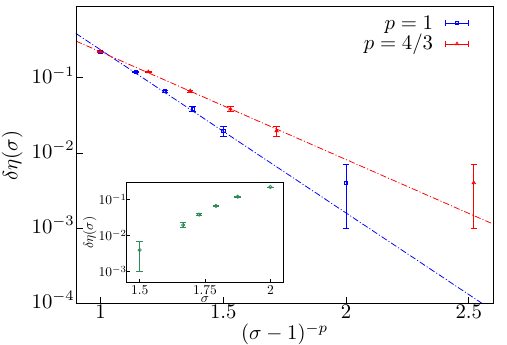}
    \caption{
    Phenomenological visualization of $\delta\eta(\sigma)=\eta(\sigma)-(2-\sigma)$. The deviation is clearly resolved in the upper part of the LR-WF-A and decreases rapidly toward the LR-WF-B interval. The reference curves illustrate the conjectured essential-singularity-like form $A\exp[-B/(\sigma-1)^p]$ for representative values of $p$.
    }
    \label{fig:delta_eta}
\end{figure}

To further illustrate the renormalization of the anomalous dimension, we plot the deviation from the GFP value, $\delta\eta=\eta-(2-\sigma)$, in the inset of Fig.~\ref{fig:delta_eta}. The deviation is clearly resolved at larger values of $\sigma$ and increases toward the LR--SR crossover. As $\sigma$ decreases, it is rapidly suppressed toward the LR-WF-B regime.

Motivated by this rapid suppression and the possibility that $\delta\eta$ approaches zero smoothly at $\sigma=1$, we use the following phenomenological essential-singularity form:
\begin{equation}
\delta\eta(\sigma)=
\begin{cases}
0, & 2/3<\sigma\le1,\\[2mm]
A\exp\!\left[-\dfrac{B}{(\sigma-1)^p}\right],
& 1<\sigma\le2,
\end{cases}
\label{eq:delta_eta_essential}
\end{equation}
where $A,B,p>0$. When extended by zero into the LR-WF-B regime, this form is infinitely differentiable at $\sigma=1$, with all derivatives vanishing at the boundary. It therefore provides a simple phenomenological description of the rapid onset of $\delta\eta$ above $\sigma=1$. As shown in Fig.~\ref{fig:delta_eta}, the data near $\sigma=1$ can be described by this form, with $p\simeq1$, although we are unable to obtain an accurate estimate of $p$.

\subsection{Coexistence of GFP and CG asymptotics in the LR-MF regime}

We now examine the LR-MF regime, where different observables exhibit distinct finite-size asymptotics. As described in Sec.~\ref{sec:two_regime_fss}, global observables that retain the leading complete-graph contribution follow CG finite-size scaling, whereas observables that filter out this contribution reveal the underlying LR-GFP scaling. We demonstrate this coexistence below for both the thermal and magnetic scaling fields.

\subsubsection{Thermal scaling}

\begin{figure}[t]
    \centering
    \includegraphics[width=1\linewidth]{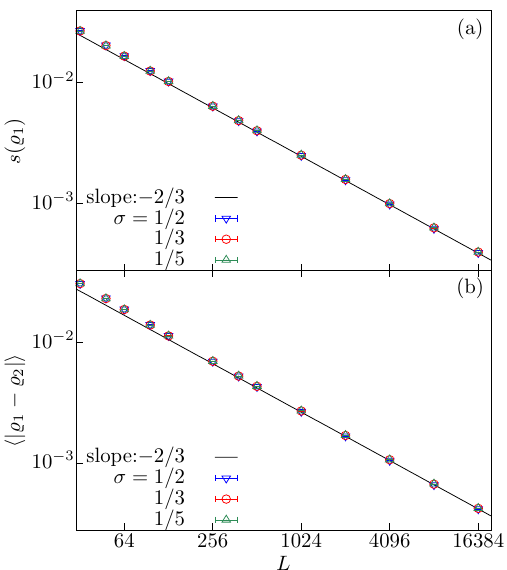}
    \caption{Log-log plot of $s(\varrho_1)$ and $\langle |\varrho_1-\varrho_2|\rangle$ as functions of system size $L$ for various $\sigma$ in the LR-MF regime. Both observables follow $L^{-2/3}$ scaling.}
    \label{fig:yt_O}
\end{figure}

We first examine the thermal scaling associated with the CG asymptotics. In the EB method, the observables $s(\varrho_1)$ and $\langle|\varrho_1-\varrho_2|\rangle$ at the pseudo-critical point are governed by the CG asymptotics, with $1/\nu_L=y_t^{\mathrm{CG}}=2/3$. 
We examine three representative values, $\sigma=1/2$, $1/3$, and $1/5$, with system sizes up to $L=16384$. 
As shown in Fig.~\ref{fig:yt_O}, both observables asymptotically follow an $L^{-2/3}$ decay at all three values of $\sigma$. We fit the two observables using the ansatz
\begin{align}
\mathcal{O} = L^{-1/\nu_{L}}(a+b_1L^{-y_1}+b_2L^{-y_2}),
\label{eq:s1_CG}
\end{align}
where $y_1$ and $y_2$ are finite-size correction exponents. 
The resulting estimates, listed in Table~\ref{table:MF}, are consistent with $y_t^{\mathrm{CG}}=2/3$ and show no systematic dependence on $\sigma$.

To determine $\rho_c$, we substitute the CG-asymptotic value $1/\nu_L=2/3$ into Eq.~\eqref{eq:rho1}. 
Because the finite-size corrections to $\rho_1$ are particularly strong in the LR-MF regime, a fit with only a single correction exponent does not yield $\chi^2/\mathrm{DF}\simeq 1$ for the accessible system sizes, where DF denotes the number of degrees of freedom. We therefore include three correction terms with exponents $\omega$, $2\omega$, and $3\omega$. We have tested several sets of correction exponents, such as $(0.1,0.2,0.3)$ and $(0.2,0.4,0.6)$. They all yield stable and mutually consistent extrapolations of $\rho_c$ within statistical uncertainty, while mainly changing the nonuniversal amplitudes $b_i$. We use $\omega=0.1$ for the estimates reported in Table~\ref{table:MF}.

We next examine the thermal scaling governed by the LR-GFP. As anticipated in Eq.~\eqref{eq:chi_k}, the Fourier transform at $\mathbf{k}\neq0$ removes the distance-independent CG contribution, leaving $\chi_{\mathbf{k}}$ governed by LR-GFP scaling. The corresponding thermal exponent is $y_t^{\mathrm{GFP}}=\sigma$.
We therefore test this prediction by plotting $(\chi_{\mathbf{k}}-c_0)/L^\sigma$ against $(\rho-\rho_c)L^\sigma$, where $c_0$ is a nonuniversal background constant.

\begin{figure*}[t]
    \centering
    \includegraphics[width=\linewidth]{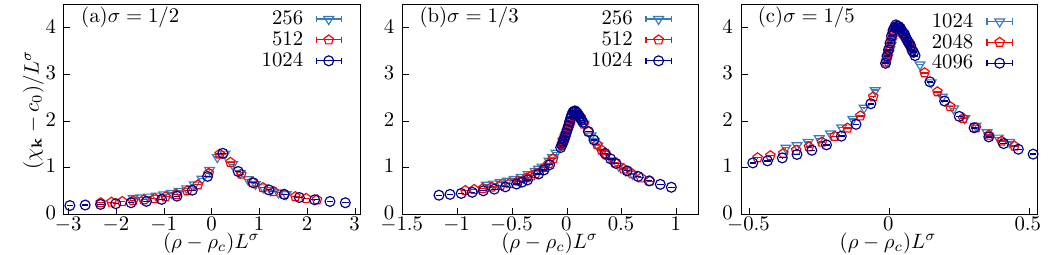}
    \caption{Data-collapse plots of $(\chi_{\mathbf{k}}-c_0)/L^{\sigma}$ versus $(\rho-\rho_c)L^{y_t^{\mathrm{GFP}}}$ for $\sigma=1/2,\,1/3,\,1/5$. For $\sigma=1/2$, $c_0=-3.1$; for $\sigma=1/3$, $c_0=-2.0$; and for $\sigma=1/5$, $c_0=-3.32$. Using the GFP prediction $y_t^{\mathrm{GFP}}=\sigma$ gives a good data collapse at all three values of $\sigma$, consistent with a GFP thermal scaling field.}
    \label{fig:MF_chik}
\end{figure*}

The resulting collapses for $\sigma=1/2$, $1/3$, and $1/5$ are shown in Fig.~\ref{fig:MF_chik}. For all three values, data for different system sizes collapse well onto a common scaling curve under the GFP scaling form
$\chi_{\mathbf{k}}-c_0
\sim
L^\sigma
\widetilde{\chi}_{\mathbf{k}}
\left[
(\rho-\rho_c)L^\sigma
\right]$. 
In particular, the scaling variable $(\rho-\rho_c)L^\sigma$ supports the GFP prediction $y_t^{\mathrm{GFP}}=\sigma$. Together with the EB results above, these results show that the LR-MF regime exhibits the CG thermal finite-size exponent $y_t^{\mathrm{CG}}=2/3$ and the LR-GFP thermal exponent $y_t^{\mathrm{GFP}}=\sigma$ in different observables.

\begin{figure}[t]
    \centering
    \includegraphics[width=1\linewidth]{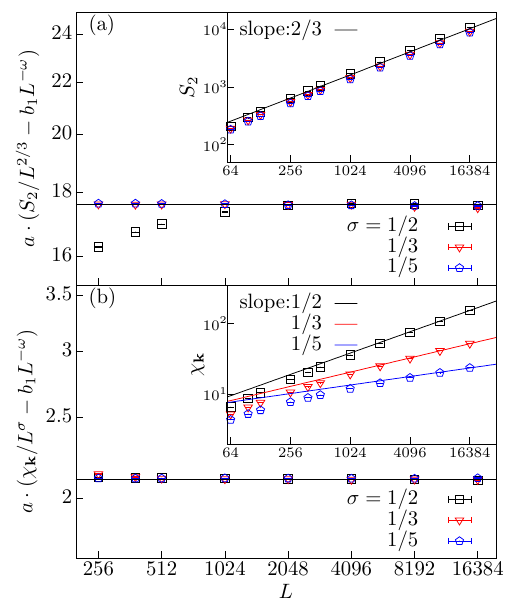}
    \caption{Log-log plots of $S_2/L^{2/3}$ and $\chi_{\mathbf{k}}/L^{\sigma}$ as functions of $L$ for $\sigma=1/2,\,1/3,\,1/5$, after subtraction of the fitted leading correction and application of a constant normalization to compare the large-$L$ behavior. For $S_2$, $\omega(1/3)=1.1617$ and $\omega(1/5)=0.8017$. At $\sigma=1/2$, the corrections do not permit a reliable estimate of $\omega$; we therefore omit the correction subtraction and show the large-$L$ behavior directly. For $\chi_{\mathbf{k}}$, $\omega(1/2)=0.4717$, $\omega(1/3)=0.6093$, and $\omega(1/5)=0.386$. After removal of the leading correction where applicable, the curves approach horizontal lines, consistent with $S_2\sim L^{2/3}$ and $\chi_{\mathbf{k}}\sim L^{\sigma}$. The insets show the raw data and the strong finite-size corrections at small $L$; solid lines are guides to the eye.}
    \label{fig:chiall}
\end{figure}

\subsubsection{Magnetic scaling}

We next turn to the magnetic scaling in the LR-MF regime. The CG and LR-GFP asymptotics again give different predictions for different observables. We fit $S_2$ and $\chi_{\mathbf{k}}$ at the pseudo-critical point using the FSS ansatz
\begin{align}
    S_2 &= L^{2-\eta_L}(a+b_1L^{-y_1}+b_2L^{-y_2})+c_0,
    \label{eq:S2_CG}
    \\
    \chi_{\mathbf{k}} &= L^{2-\eta^{\mathrm{GFP}}}(a+b_1L^{-y_1}+b_2L^{-y_2})+c_0,
    \label{eq:chik_GFP}
\end{align}
where $y_1$ and $y_2$ are finite-size correction exponents, $a$, $b_1$, and $b_2$ are nonuniversal constants, and $c_0$ is a background term. The fitting methodology and detailed results are given in Appendix~\ref{sec:fitting_details}; the final estimates are listed in Table~\ref{table:MF}.

Fig.~\ref{fig:chiall} compares the two magnetic scaling behaviors. After division by the predicted leading power laws and subtraction of the fitted leading correction where applicable, both rescaled quantities approach horizontal lines at large $L$. The insets show the raw Monte Carlo data. Despite strong finite-size corrections, $S_2$ and $\chi_{\mathbf{k}}$ exhibit clearly distinct asymptotic scaling: $S_2\sim L^{2/3}$ follows the CG prediction, whereas $\chi_{\mathbf{k}}\sim L^\sigma$ follows the LR-GFP prediction.

The GFP relation $\eta^{\mathrm{GFP}}=2-\sigma$ is also consistent with the CE collapse in Fig.~\ref{fig:MF_chik}. In that collapse, the horizontal scaling variable $(\rho-\rho_c)L^\sigma$ tests the thermal exponent $y_t^{\mathrm{GFP}}=\sigma$, whereas the vertical rescaling $(\chi_{\mathbf{k}}-c_0)/L^\sigma$ follows the magnetic prediction $2-\eta^{\mathrm{GFP}}=\sigma$. Thus, the same CE data are simultaneously consistent with the thermal and magnetic GFP scaling forms.

To summarize, the LR-MF regime exhibits coexisting CG finite-size asymptotics and LR-GFP scaling. The EB observables yield $y_t^{\mathrm{CG}}=2/3$, while the CE collapse follows $y_t^{\mathrm{GFP}}=\sigma$ (Table~\ref{table:MF}). For the magnetic observables, $S_2$, $C_1$, and $\Delta$ follow the CG prediction $y_h^{\mathrm{CG}}=4/3$, whereas $\chi_{\mathbf{k}}$ follows the GFP prediction $y_h^{\mathrm{GFP}}=1+\sigma/2$. 
The simultaneous observation of these two asymptotic behaviors across independent observables and simulation methods supports the coexistence of CG finite-size asymptotics and LR-GFP scaling in the LR-MF regime.

\begin{table*}[t]
\centering
\small
\setlength{\tabcolsep}{3pt}
\renewcommand{\arraystretch}{1.15}
\caption{
Summary of the proposed scaling scenario for 2D LR-percolation. The columns specify the regimes in $\sigma$, the corresponding LR-SRW behavior, the thermodynamic exponents $(\eta,\nu)$, and the finite-size scaling exponents $(\eta_L,\nu_L)$. For reference, the SR Gaussian exponents are $(\eta_{\mathrm{SR}}^{\mathrm{GFP}},\nu_{\mathrm{SR}}^{\mathrm{GFP}})=(0,1/2)$, the LR Gaussian exponents are $(\eta^{\mathrm{GFP}},\nu^{\mathrm{GFP}})=(2-\sigma,1/\sigma)$, and the complete-graph finite-size exponents in $d=2$ are $(\eta_L,\nu_L)=(4/3,3/2)$.
}
\label{table:universality}
\begin{tabular*}{\textwidth}{@{\extracolsep{\fill}}c|llll}
\hline\hline
&
\begin{tabular}[t]{@{}l@{}}SR-WF\\ $\sigma>2$\end{tabular}
&
\begin{tabular}[t]{@{}l@{}}LR-WF-A\\ $1<\sigma\le2$\end{tabular}
&
\begin{tabular}[t]{@{}l@{}}LR-WF-B\\ $2/3<\sigma\le1$\end{tabular}
&
\begin{tabular}[t]{@{}l@{}}LR-MF\\ $0<\sigma\le2/3$\end{tabular}
\\
\hline
LR-SRW
& diffusive
& sub-ballistic
& hyper-ballistic
& hyper-ballistic
\\
\hline
$\eta$
& $\eta_{\mathrm{SR}}=5/24$
& $\eta=2-\sigma+\delta\eta(\sigma)$
& $\eta=2-\sigma$
& $\eta=2-\sigma$
\\
$\nu$
& $\nu_{\mathrm{SR}}=4/3$
& non-Gaussian
& non-Gaussian
& $\nu=1/\sigma$
\\
$\eta_L$
 & $\eta_L=\eta_{\mathrm{SR}}$
& $\eta_L=\eta$
& $\eta_L=2-\sigma$
& $\eta_L=4/3$
\\
$\nu_L$
& $\nu_L=\nu_{\mathrm{SR}}$
& $\nu_L=\nu$
& $\nu_L=\nu$
& $\nu_L=3/2$
\\
\hline\hline
\end{tabular*}
\end{table*}

\section{Discussion}
\label{sec:conclusion}

We studied the critical behavior of 2D LR bond percolation with connection probability $p(r)\propto 1/r^{2+\sigma}$.
Combining event-based and conventional ensembles, with system sizes up to $L=16384$, we analyzed the FSS behavior on the long-range side of the LR--SR crossover.
Together with the preceding evidence that the LR--SR boundary is located at $\sigma^*=2$~\cite{liu2025twodimensionalpercolationmodellongrange}, the present results are consistent with the proposed universality picture for the 2D LR-percolation model: SR-WF for $\sigma>2$, LR-WF-A for $1<\sigma\le2$, LR-WF-B for $2/3<\sigma\le1$, and LR-MF for $0<\sigma\le2/3$.
Table~\ref{table:universality} summarizes the exponents in these regimes and includes the transport properties of the LR-SRW for comparison.

In the LR-WF-B regime, $2/3<\sigma\le1$, the anomalous dimension is consistent with
\begin{equation}
\eta=2-\sigma .
\end{equation}
At the same time, the correlation-length exponent is clearly not described by the Gaussian prediction $\nu=1/\sigma$.
Thus, LR-WF-B is an interacting LR regime in which the anomalous dimension retains its Gaussian-form value while the thermal scaling field remains nontrivial.

In the LR-WF-A regime, $1<\sigma\le2$, the data reveal a different trend.
For $1<\sigma\lesssim 3/2$, the anomalous dimension remains compatible with $\eta=2-\sigma$ within the present numerical resolution.
For larger $\sigma$, especially for $\sigma\gtrsim3/2$, the rescaled magnetic observables develop visible large-$L$ slopes, and the fitted values of $\eta$ show statistically resolvable deviations from the simple form $2-\sigma$.
The thermal exponent also varies nontrivially with $\sigma$ and does not follow the Gaussian value.
These results are consistent with the upper part of the LR-WF-A regime being governed by an interacting LR-WF fixed point rather than by a direct continuation of either the LR Gaussian fixed point or the SR-WF fixed point. 
These results are consistent with the recent $6-\epsilon$ perturbative RG study of the LR-percolation model~\cite{li20266epsilonexpansionlongrangeleeyang}. More broadly, the structure of the LR--SR crossover boundary is also in line with that found in LR-O$(n)$ spin models~\cite{li4eExpansionLongrange2026}, suggesting a more general characterization of this boundary.

The visible onset near $\sigma\simeq3/2$ marks the scale at which the deviation of $\eta$ becomes resolvable in the present finite-size data, rather than a sharp phase boundary. The proposed boundary at $\sigma=1$ is instead motivated by the change of transport geometry in the corresponding LR-SRW, and by the broader universality picture~\cite{xiao2025universalitydiagramphasetransitions}.
A theoretical understanding of how the correction $\eta-(2-\sigma)$ emerges from the LR-WF-A regime, and how it vanishes toward the LR-WF-B side, remains an important open problem.

In the LR-MF regime, we find evidence for the coexistence of CG and GFP FSS behavior. Global pseudo-critical observables follow the CG-asymptotic exponents, whereas the nonzero-momentum susceptibility follows the GFP exponents. This coexistence is analogous to the finite-size scaling structure of systems above their upper critical dimensions under periodic boundary conditions~\cite{PhysRevE.110.044140,PhysRevE.102.022125}. At the marginal point $\sigma=2/3$, our estimates are compatible with the leading mean-field powers, but possible multiplicative logarithmic corrections are not analyzed here.

\acknowledgments
We acknowledge the support from the National Natural Science Foundation of China (NSFC) under Grant No. 12204173 and No. 12275263, as well as Quantum Science and Technology-National Science and Technology Major Project (under Grant No. 2021ZD0301900). YD is also supported by the Natural Science Foundation of Fujian Province, China (Grant No. 2023J02032). ZF is also supported by the National Natural Science Foundation of China (NSFC) under Grant No. 12504265. 

\appendix

\section{Fitting details for thermal and magnetic observables}
\label{sec:fitting_details}

This appendix collects the detailed fitting results for the observables used to extract the critical exponents in the LR-WF-A, LR-WF-B, and LR-MF regimes. All fits are performed using the standard least-squares method with a lower cutoff $L\ge L_{\min}$; the preferred fit is chosen such that $\chi^2/\mathrm{DF}\approx1$, and systematic uncertainties are estimated by comparing fits with different reasonable correction forms.

\subsection{Thermal observables}

For each $\sigma$, we fit $s(\varrho_1)$ and $\langle|\varrho_1-\varrho_2|\rangle$ using the FSS ansatz Eq.~\eqref{eq:s1}.
As an illustrative example, we consider the fit of $s(\varrho_1)$ at $\sigma = 19/11$. We first include only one correction term $y_1$ and perform an unconstrained least-squares fit with a cutoff $L_{\min}=24$, to obtain results with $\chi^2/\mathrm{DF}$ as close to $1$ as possible. In practice, this procedure consists of systematically increasing $L_{\min}$ and monitoring the variation of $\chi^2/\mathrm{DF}$, so that the preferred fit is taken as the smallest $L_{\min}$ that yields an acceptable goodness of fit. Simultaneously, we include a subleading correction term, $y_2$, and fix it to $1$ or $2$ to assess its impact on the fit; the resulting estimates are in close agreement with those obtained when only the single correction $y_1$ is included. Combining the fits obtained with different correction terms, we obtain an estimate for $1/\nu$. The fitting procedures for all thermal and magnetic observables at different values of $\sigma$ follow a similar approach, and the fitting details of $s(\varrho_1)$ in the LR-WF-A and LR-WF-B regimes are presented in Table~\ref{table:sig1}.

\begin{table*}
\caption{Fits of $s(\varrho_1)$ for selected values of $\sigma$ in the LR-WF-A and LR-WF-B regimes.}\label{table:sig1}
\begin{tabular*}{\textwidth}{@{\extracolsep{\fill}}llllllllll}
\hline\hline
$\sigma$ & $L_{\min}$ & $\chi^2/\mathrm{DF}$ & $1/\nu$       & $a$         & $y_1$   & $y_2$ & $b_1$      & $b_2$    \\ \hline
19/11    & 24         & 9.4/10      & 0.715\,2(4) & 0.406(2)    & 0.58(2) & -     & -0.215(7)  & -        \\
         & 32         & 9.3/9       & 0.715\,3(5) & 0.406(2)    & 0.58(3) & -     & -0.21(1)   & -        \\
         & 12         & 9.5/11      & 0.715\,4(7) & 0.406(3)    & 0.56(6) & 1     & -0.20(4)   & -0.02(7) \\
         & 8          & 17.5/12     & 0.716\,2(6) & 0.410(2)    & 0.52(2) & 2     & -0.189(6)  & -0.35(6) \\
         & 12         & 9.5/11      & 0.715\,3(5) & 0.406(2)    & 0.57(3) & 2     & -0.21(1)   & -0.0(1)  \\ \hline
5/3      & 64         & 4.7/7       & 0.712\,2(6) & 0.395(2)    & 0.59(5) & -     & -0.19(2)   & -        \\
         & 96         & 2.3/6       & 0.713\,4(9) & 0.401(4)    & 0.47(6) & -     & -0.13(2)   & -        \\
         & 32         & 5.1/8       & 0.712(1)    & 0.396(4)    & 0.6(2)  & 1     & -0.2(2)    & \,0.1(3) \\
         & 12         & 9.2/11      & 0.712\,8(5) & 0.398(2)    & 0.53(3) & 2     & -0.157(8)  & -0.5(1)  \\ \hline
3/2      & 24         & 7.0/10      & 0.708\,0(2) & 0.371\,8(7) & 0.83(3) & -     & -0.20(2)   & -        \\
         & 32         & 7.0/9       & 0.708\,0(3) & 0.372\,0(9) & 0.82(5) & -     & -0.19(2)   & -        \\
         & 12         & 7.8/11      & 0.708\,2(6) & 0.373(2)    & 0.6(3)  & 1     & -0.05(6)   & -0.18(8) \\
         & 8          & 7.8/12      & 0.708\,2(2) & 0.372\,6(7) & 0.76(3) & 2     & -0.16(1)   & -0.29(6) \\
         & 12         & 7.4/11      & 0.708\,1(3) & 0.372\,2(9) & 0.79(5) & 2     & -0.18(2)   & -0.2(2)  \\ \hline
4/3      & 24         & 12.3/9      & 0.710\,6(4) & 0.364(1)    & 0.83(8) & -     & -0.13(2)   & -        \\
         & 32         & 9.7/8       & 0.711\,0(5) & 0.365(2)    & 0.7(1)  & -     & -0.10(2)   & -        \\
         & 8          & 10.3/11     & 0.711\,3(5) & 0.367(2)    & 0.61(6) & 2     & -0.070(8)  & -0.69(5) \\
         & 12         & 9.8/10      & 0.711\,1(5) & 0.366(2)    & 0.7(1)  & 2     & -0.08(2)   & -0.6(2)  \\ \hline
6/5      & 24         & 11.1/8      & 0.710\,7(2) & 0.352\,8(6) & 1.3(2)  & -     & -0.2(1)    & -        \\
         & 32         & 10.5/7      & 0.710\,9(4) & 0.353(1)    & 1.0(3)  & -     & -0.08(8)   & -        \\
         & 8          & 12.6/10     & 0.710\,6(2) & 0.352\,6(6) & 1.7(2)  & 1     & -0.5(1)    & -0.03(4) \\
         & 12         & 10.4/9      & 0.710\,7(2) & 0.352\,9(5) & 2.3(6)  & 1     & -1(2)      & -0.06(3) \\ \hline
1        & 32         & 10.5/8      & 0.704(1)    & 0.324(5)    & 0.4(1)  & -     & \,0.042(4) & -        \\
         & 64         & 7.1/6       & 0.705(1)    & 0.328(4)    & 0.6(2)  & -     & \,0.05(3)  & -        \\
         & 8          & 13.3/11     & 0.704\,9(7) & 0.326(2)    & 0.53(7) & 2     & \,0.059(6) & -1.04(5) \\
         & 12         & 12.0/10     & 0.705\,3(7) & 0.327(2)    & 0.6(1)  & 2     & \,0.07(1)  & -1.2(1)  \\
         & 24         & 8.4/8       & 0.706\,1(6) & 0.330(2)    & 0.8(2)  & 2     & \,0.1(1)   & -2(1)    \\
         & 32         & 8.4/7       & 0.706\,1(7) & 0.330(2)    & 0.9(3)  & 2     & \,0.2(2)   & -2(2)    \\ \hline
9/10     & 64         & 4.7/6       & 0.699\,4(7) & 0.312(2)    & 0.68(9) & -     & \,0.14(4)  & -        \\
         & 96         & 4.2/5       & 0.699(1)    & 0.310(4)    & 0.6(2)  & -     & \,0.10(5)  & -        \\
         & 12         & 11.9/10     & 0.699\,1(5) & 0.311(1)    & 0.96(9) & 1     & \,5(1)     & -5(1)    \\
         & 24         & 9.4/8       & 0.699\,2(8) & 0.311(2)    & 0.9(3)  & 1     & \,2(1)     & -2(1)    \\
         & 12         & 15.0/10     & 0.698\,2(9) & 0.308(3)    & 0.56(6) & 2     & \,0.11(1)  & -1.2(1)  \\
         & 24         & 9.3/8       & 0.699\,1(8) & 0.311(2)    & 0.7(1)  & 2     & \,0.15(5)  & -1.8(8)  \\ \hline
5/6      & 64         & 4.0/6       & 0.693\,2(8) & 0.297(2)    & 0.60(6) & -     & \,0.15(2)  & -        \\
         & 96         & 3.9/5       & 0.694(1)    & 0.298(3)    & 0.6(1)  & -     & \,0.17(6)  & -        \\
         & 32         & 5.6/7       & 0.693(1)    & 0.297(4)    & 0.7(2)  & 1     & \,0.3(5)   & -0.4(6)  \\
         & 24         & 5.0/8       & 0.693\,2(7) & 0.297(2)    & 0.62(6) & 2     & \,0.16(3)  & -1.7(5)  \\
         & 32         & 4.6/7       & 0.693\,5(8) & 0.298(2)    & 0.66(9) & 2     & \,0.19(5)  & -2(1)    \\ \hline
3/4      & 96         & 5.6/5       & 0.682(2)    & 0.272(7)    & 0.48(9) & -     & \,0.15(3)  & -        \\
         & 128        & 5.5/4       & 0.683(3)    & 0.273(8)    & 0.5(1)  & -     & \,0.16(5)  & -        \\
         & 24         & 6.1/8       & 0.683(2)    & 0.274(5)    & 0.5(1)  & 1     & \,0.20(8)  & -0.1(1)  \\
         & 12         & 10.9/10     & 0.683\,0(9) & 0.275(2)    & 0.52(3) & 2     & \,0.168(8) & -1.2(1)  \\
         & 24         & 6.2/8       & 0.682(1)    & 0.271(4)    & 0.47(5) & 2     & \,0.15(1)  & -0.5(4)  \\ \hline\hline
\end{tabular*}
\end{table*}

\subsection{Magnetic observables}

\begin{table*}
\caption{Fits of $S_2$ for $\sigma=1/2$, $1/3$, and $1/5$.}\label{table:MF_S2}
\begin{tabular*}{\textwidth}{@{\extracolsep{\fill}}llllllllll}
\hline\hline
$\sigma$ & $L_{\min}$ & $\chi^2/\mathrm{DF}$ & $\eta_L$    & $a$      & $y_1$   & $b_1$     & $c_0$   \\ \hline
1/2      & 256        & 3.9/4       & 4/3       & 10(2)    & 0.03(6) & 9(2)      & -88(3)  \\ 
        & 256         & 3.2/5       & 1.3443(3)  & 18.98(4) & - & - & -87.6(7) \\
         & 384 & 0.7/4 & 1.3447(2)  & 19.07(3) & - & -  & -89.3(6) \\ \hline
1/3      & 384        & 7.6/3       & 1.339(3)  & 15.2(4)  & 0.8(2)  & -92(91)   & -       \\
         & 512        & 7.3/2       & 1.340(5)  & 15.4(8)  & 0.7(4)  & -61(112)  & -       \\ \hline
1/5      & 512        & 3.2/2       & 1.333(2)  & 13.6(2)  & 1.0(3)  & -203(288) & -       \\
         & 1024       & 1.9/2       & 1.3344(9) & 13.9(1)  & -       & -         & -27(4)  \\
\hline\hline
\end{tabular*}
\end{table*}

In the LR-WF-A and LR-WF-B regimes, $S_2$ is fitted using Eq.~\eqref{eq:S2_scaling}, whereas $C_1$ and $\Delta$ are fitted using Eq.~\eqref{eq:Delta_scaling}. In the LR-MF regime, $S_2$ and $\chi_{\mathbf{k}}$ are fitted using Eqs.~\eqref{eq:S2_CG} and~\eqref{eq:chik_GFP}, respectively. The fitting strategy is the same as that used for the thermal observables. Detailed fits of $\Delta$ in the LR-WF-A and LR-WF-B regimes are given in Table~\ref{table:Delta}; fits of $\chi_{\mathbf{k}}$ and $S_2$ in the LR-MF regime are given in Tables~\ref{table:MF_chik} and~\ref{table:MF_S2}.

\begin{table*}
\caption{Fits of $\Delta$ for selected values of $\sigma$ in the LR-WF-A and LR-WF-B regimes.}\label{table:Delta}
\begin{tabular*}{\textwidth}{@{\extracolsep{\fill}}llllllllll}
\hline\hline
$\sigma$ & $L_{\min}$ & $\chi^2/\mathrm{DF}$ & $\eta$    & $a$      & $y_1$     & $y_2$ & $b_1$    & $b_2$ & $c_0$ \\ \hline
19/11    & 128        & 5.7/5       & 0.313(2)  & 0.193(3)  & 0.41(2)  & -     & 0.181(7)  & -     & -     \\
         & 256        & 5.4/4       & 0.311(5)  & 0.191(6)  & 0.38(5)  & -     & 0.17(2)   & -     & -     \\
         & 64         & 5.4/6       & 0.311(3) & 0.190(4) & 0.38(2) & 2     & 0.169(7) & 3(1)     & -     \\
         & 96         & 5.4/5       & 0.311(4) & 0.191(5) & 0.38(4) & 2     & 0.17(2)  & 2(3)     & -     \\
         & 48         & 5.6/7       & 0.311(2) & 0.191(3) & 0.38(2) & -     & 0.170(6) & -     & 1.5(3)   \\  \hline
5/3      & 128        & 1.5/5       & 0.355(1)  & 0.206(2)  & 0.40(1)  & -     & 0.177(4)  & -     & -     \\
         & 256        & 1.2/4       & 0.353(2)  & 0.204(3)  & 0.38(2)  & -     & 0.167(8)  & -     & -     \\
         & 24         & 10.8/9      & 0.353(3) & 0.203(4) & 0.35(3) & 1     & 0.144(7) & 0.16(3)  & -     \\
         & 64         & 1.4/6       & 0.353(1) & 0.204(2) & 0.38(1) & 2     & 0.166(4) & 2.6(5)   & -     \\
         & 64         & 1.4/6       & 0.352(2) & 0.203(2) & 0.37(1) & -     & 0.164(4) & -     & 1.5(3)   \\ \hline
3/2      & 64         & 3.5/7       & 0.504 7(7) & 0.277(1)  & 0.52(1)  & -     & 0.184(6) & -     & -     \\
         & 96         & 3.3/6       & 0.505 0(9) & 0.278(1)  & 0.53(2)  & -     & 0.19(1)  & -     & -     \\
         & 128        & 3.3/5       & 0.505(1)  & 0.278(2)  & 0.54(3)  & -     & 0.19(2)  & -     & -     \\
         & 32         & 4.4/8       & 0.505(1)  & 0.278(2)  & 0.57(5) & 1     & 0.23(5) & -0.10(8) & -     \\
         & 24         & 4.8/9       & 0.504 5(7) & 0.277(1)  & 0.52(1)  & 2     & 0.183(6) & -0.3(2)  & -     \\
         & 8          & 7.4/12      & 0.504 3(4) & 0.276 9(6) & 0.522(6) & -     & 0.185(2) & -     & -0.19(1) \\ \hline
4/3      & 48         & 5.4/7       & 0.664(1) & 0.361(2) & 0.45(4) & -     & 0.086(4)  & -     & -     \\
         & 16         & 9.1/9       & 0.666 8(8) & 0.367(1) & 0.75(7) & 1     & 0.4(2)  & -0.5(2)  & -     \\
         & 24         & 7.1/8       & 0.665(1)  & 0.365(3) & 0.6(1)  & 1     & 0.17(9) & -0.2(1)  & -     \\
         & 16         & 6.7/9       & 0.664 7(9) & 0.363(2) & 0.50(3) & 2     & 0.097(5) & -0.7(1)  & -     \\
         & 16         & 7.3/9       & 0.665 4(9) & 0.365(2)      & 0.54(4) & -     & 0.110(9) & -     & -0.39(6) \\
         & 24         & 6.6/8       & 0.665(1)  & 0.363(3)      & 0.50(6) & -     & 0.10(1)  & -     & -0.3(1)  \\ \hline
6/5      & 24         & 9.9/8       & 0.800 6(2) & 0.458 1(3) & 1.8(2)  & -     & -1.1(6)  & -     & -     \\  
         & 96         & 8.3/5       & 0.800 5(3) & 0.457 8(4) & -     & -     & -     & -     & -0.1(3)  \\
         & 512        & 7.3/3       & 0.800 4(4) & 0.457 6(6) & -     & -     & -     & -     & -     \\
         & 1024       & 2.6/2       & 0.799 9(4) & 0.456 7(7) & -     & -     & -     & -     & -     \\ \hline
1        & 48         & 6.0/7       & 1.000 3(5) & 0.634(1) & 0.77(2) & -     & -0.56(4)  & -     & -     \\
         & 64         & 5.9/6       & 1.000 4(7) & 0.634(2) & 0.76(4) & -     & -0.54(6)  & -     & -     \\
         & 16         & 9.3/9       & 1.000 3(5) & 0.634(1) & 0.79(2) & 2     & -0.60(4) & 1.6(3)  & -     \\
         & 24         & 6.3/8       & 0.999 6(5) & 0.632(1) & 0.84(3) & 2     & -0.73(9) & 3.0(8)  & -     \\
         & 12         & 10.3/10     & 1.000 2(5) & 0.634(1) & 0.81(2) & -     & -0.70(6) & -     & 0.8(1)  \\ \hline
9/10     & 64         & 1.3/6       & 1.100 2(4) & 0.756(1) & 0.68(1) & -     & -0.78(2)  & -     & -     \\
         & 8          & 11.1/11     & 1.100 0(5) & 0.755(2) & 0.78(2) & 1     & -1.8(2) & 1.5(2)  & -     \\
         & 16         & 11.8/9      & 1.100 6(8) & 0.757(3) & 0.69(2) & -     & -0.86(6) & -     & 0.8(2)  \\ \hline
5/6      & 48         & 8.7/7       & 1.168(1)  & 0.866(4) & 0.55(1)  & -     & -0.78(2)  & -     & -     \\
         & 64         & 5.4/6       & 1.167(1)  & 0.861(4) & 0.57(2)  & -     & -0.83(3)  & -     & -     \\
         & 16         & 13.2/9      & 1.166 6(9) & 0.860(4) & 0.58(1) & 2     & -0.85(3) & 1.9(3) & -     \\
         & 24         & 13.4/8      & 1.165(1)  & 0.855(6) & 0.62(3) & -     & -1.0(1)  & -     & 0.9(4)  \\ \hline
3/4      & 128        & 11.9/4      & 1.245(5) & 1.01(3)  & 0.48(6) & -     & -0.9(1)   & -     & -     \\
         & 48         & 12.9/6      & 1.245(6)  & 1.01(4)  & 0.5(1)  & 1     & -0.8(3)  & -0.3(8) & -     \\
         & 64         & 12.5/5      & 1.246(9)  & 1.02(6)  & 0.4(2)  & 1     & -0.7(4)  & -1(1)   & -     \\
         & 48         & 12.9/6      & 1.244(5)  & 1.01(3)  & 0.47(7)  & -     & -0.9(2)  & -     & -1(1)   \\
\hline\hline
\end{tabular*}
\end{table*}

\begin{table*}
\caption{Fits of $\chi_{\mathbf{k}}$ for $\sigma=1/2$, $1/3$, and $1/5$.}\label{table:MF_chik}
\begin{tabular*}{\textwidth}{@{\extracolsep{\fill}}llllllllll}
\hline\hline
$\sigma$ & $L_{\min}$ & $\chi^2/\mathrm{DF}$ & $\eta^{\mathrm{GFP}}$    & $a$     & $y_1$   & $y_2$ & $b_1$    & $b_2$  & $c_0$    \\ \hline
1/2      & 384        & 2.0/3       & 1.507(4)  & 1.33(7) & 0.42(5) & -     & -2.8(4)  & -      & -        \\
         & 512        & 1.2/2       & 1.512(8)  & 1.4(2)  & 0.35(8) & -     & -2.4(3)  & -      & -        \\
         & 512        & 3.5/3       & 1.501 0(8) & 1.24(1) & -       & -     & -        & -      & -3.54(8) \\
         & 1024       & 2.6/2       & 1.502(2)  & 1.26(2) & -       & -     & -        & -      & -3.7(2)  \\
         & 96         & 3.3/5       & 1.510(8)  & 1.4(2)  & 0.4(3)  & -     & -6(3)    & -      & \,4(3)   \\
         & 96         & 2.4/5       & 1.506(3)  & 1.32(6) & 0.44(5) & 1     & -3.2(6)  & 3(2)   & -        \\
         & 128        & 2.3/4       & 1.508(4)  & 1.35(6) & 0.40(4) & 2     & -2.7(3)  & 64(7)  & -        \\  \hline
1/3      & 512        & 2.6/2       & 1.670(6)  & 2.2(2)  & 0.5(1)  & -     & -10(3)   & -      & -        \\
         & 384        & 3.9/2       & 1.666(9)  & 2.1(2)  & 0.9(6)  & 1     & -270(2)  & 381(3) & -        \\
         \hline
1/5      & 256        & 5.3/4       & 1.799(4)  & 3.6(2)  & 0.39(3) & -     & -8.9(5)  & -      & -        \\
         & 384        & 2.8/3       & 1.793(4)  & 3.3(2)  & 0.46(5) & -     & -10(1)   & -      & -        \\
         & 32         & 16.0/8      & 1.799(3)  & 3.5(1)  & 0.43(2) & 1     & -11.5(7) & 17(2)  & -          \\
\hline\hline
\end{tabular*}
\end{table*}

\clearpage
\bibliography{ref}
\end{document}